\documentclass[conference,letter,10pt]{IEEEtran}
\usepackage{times}
\usepackage{amsfonts}
\usepackage{color}
\usepackage[dvips]{graphicx}
\usepackage[cmex10]{amsmath}
\usepackage[numbers,sort&compress]{natbib}
\usepackage[tight,footnotesize]{subfigure}

\renewcommand{\baselinestretch}{1.02}

\title{Antenna Array Signal  Processing for Quaternion-Valued Wireless Communication Systems}

\author{Wei Liu\vspace*{0.2cm}\\

      {\normalsize Communications Research Group}\\
      {\normalsize Dept. of Electronic \& Electrical Engineering}\\
        {\normalsize University of Sheffield, Sheffield S1 3JD, UK}\\
         { \tt w.liu@sheffield.ac.uk }}

\begin{document}
\maketitle
\begin{abstract}
Quaternion-valued wireless communication systems have been studied in the past. Although progress has been made in this promising area, a crucial missing link  is lack of effective and efficient quaternion-valued signal processing algorithms for channel equalisation and beamforming. With most recent developments in quaternion-valued signal processing, in this work, we fill the gap to solve the  problem and further derive the quaternion-valued Wiener solution for block-based calculation.
\end{abstract}

\section{Introduction}
\label{sec:introdcution}

Increasing the capacity of a wireless communication system has always been a focus of the wireless communications research community. It is well-known that polarisation diversity can be exploited to mitigate the multipath effect to maintain a reliable communication link with an acceptable quality of service (QoS), where a pair of antennas with orthogonal polaristion directions is employed at both the transmitter and the receiver sides. However, the traditional diversity scheme aims to achieve a single reliable channel link between the transmitter and the receiver, while the same information is transmitted at the same frequency but with different polarisations, i.e. two channels. This is not an effective use of the precious spectrum resources as the two channels could be used to transmit different data streams simultaneously. For example, we can design a four-dimensional (4-D) modulation scheme across the two polarisation diversity channels using a quaternion-valued representation, as proposed in \cite{isaeva95a}. An earlier version of quaternion-valued 4-D modulation scheme based on two different frequencies was proposed in \cite{zetterberg77a}. However, due to the change of polarisation of the transmitted radio frequency signals during the complicated propagation process including multipath, reflection, refraction, etc, interference will be caused to each other at the two differently polarised receiving antennas. To solve the problem, efficient signal processing methods and algorithms for channel equalisation and interference suppression/beamforming are needed for practical implementation of the proposed 4-D modulation scheme.

Recently, quaternion-valued signal processing has been introduced and studied in details to solve  problems related to three or four-dimensional signals \cite{bihan04a}, such as vector-sensor array signal processing~\cite{liu14e}, and wind profile prediction~\cite{liu14f}. With most recent developments in this area, especially the derivation of quaternion-valued gradient operators and the quaternion-valued least mean square (QLMS) algorithm \cite{liu14f,liu14p}, we are now ready to effectively solve the 4-D equalisation and interference suppression/beamforming problem associated with the proposed  4-D modulation scheme. Now the dual-channel effect on the transmitted signal can be modelled by a quaternion-valued IIR/FIR filter. At the receiver side, for channel equalisation, we can employ a quaternion-valued adaptive algorithm to recover the original 4-D signal, which inherently also performs an interference suppression operation to separate the original two 2-D signals. Moreover, multiple antenna pairs can be employed at the receiver side to perform the traditional beamforming task to suppress other interfering signals. Note although quaternion-valued wireless communication employing multiple antennas has been studied before, such as the design of orthogonal space-time-polarization block code in \cite{wysocki09a}, to our best knowledge, it is the first time to study the quaternion-valued equalisation and interference suppression/beamforming problem in this context.

In the following, the 4-D modulation scheme based on two orthogonally polarised antennas will be introduced in Sec. \ref{sec:modulation} and the required quaternion-valued equalisation and inter-channel interference suppression solution and their extension to multiple dual-polarised antennas are presented in Sec. \ref{sec:equal_beam}. Simulation results are provided in Sec. \ref{sec:sim}, followed by conclusions in Sec. \ref{sec:concl}.


\section{Quaternion-valued 4-D Modulation}
\label{sec:modulation}

\begin{figure}
	\centering
	\includegraphics[scale=0.69]{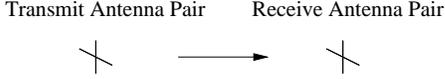}
	\caption{Wireless communication employing polarisation diversity, where both the transmitter and the receiver sides are equipped with a pair of antennas with different polarisation directions.}
	\label{fig:TRantennapair}
\end{figure}

In traditional polarisation diversity scheme, as shown in Fig.~\ref{fig:TRantennapair},
each side is equipped with two antennas with orthogonal polarisation directions and the signal being transmitted is two-dimensional, i.e. complex-valued with one real part and one imaginary part. In the quaternion-valued modulation scheme, the signal is modulated across the two antennas to generate a 4-D modulated signal. Such a signal can be conveniently represented mathematically by a quaternion \cite{hamilton66a,kantor89a}.
A quaternion is a hypercomplex number defined as
\begin{equation}
\label{eq:quaternion}
    q=q_0 + iq_1 + jq_2 + kq_3\;,
\end{equation}
where $q_0$ is the real part of the quaternion, and $q_1$, $q_2$ and
$q_3$ are the three imaginary components with their corresponding imaginary units $i$, $j$ and $k$, respectively.

As an example, corresponding to the 4-QAM (Quadrature Amplitude Modulation) in the two-dimensional case, for the 4-D modulation scheme, $q_0$, $q_1$, $q_2$ and $q_3$ can take values of either $1$ or $-1$, representing $16$ different symbols. We can call this scheme 16-QQAM (Quaternion-valued QAM) or 16-$\mbox{Q}^2\mbox{AM}$.

\section{Quaternion-valued equalisation and interference suppression/beamforming}
\label{sec:equal_beam}
\begin{figure}
	\centering
	\includegraphics[scale=0.49]{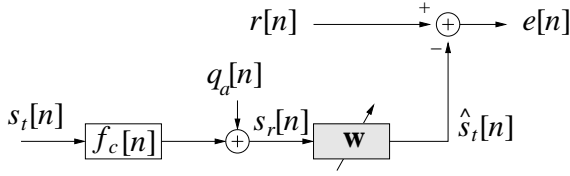}
	\caption{Channel model and the reference signal based equaliser for quaternion-valued signals}
	\label{fig:channelmodel}
\end{figure}

The signal transmitted by the two antennas will go through the channel with all kinds of effects and arrive at the receiver side, where the two antennas with orthogonal polarisation directions (Note the orthogonal polarisation may not give the best performance) will pick up the two signals. Again the four components of the received signal can be represented by another quaternion. We use $s_t[n]$ and $s_r[n]$ to represent the transmitted and received 4-D quaternion-valued signals, respectively. Then the channel effect can be modeled by a filter with quaternion-valued impulse response $f_c[n]$, i.e.
\begin{equation}
\label{eq:channelmodel}
    s_r[n]=s_t[n]*f_{c}[n]+q_a[n]\;,
\end{equation}
where $q_a[n]$ is the quaternion-valued additive noise, as shown in Fig.~\ref{fig:channelmodel}.

To recover $s_t[n]$ from $s_r[n]$ or estimate the channel, as in the 2-D case (complex-valued), we can design a quaternion-valued equaliser. One choice is a reference signal based equaliser, among many others corresponding to the complex-valued case. Now assume we have a reference signal $r[n]$ available, then we can employ the standard adaptive filter structure shown in the second half of Fig.~\ref{fig:channelmodel} and update the equaliser coefficient vector $\mathbf{w}$ with a length of $L$ by minimising the mean square value of the error signal $e[n]$~\cite{liu14f,liu14p}.

The cost function is given by
\begin{equation}
J[n]=e[n]e^*[n]\;,
\end{equation}
where
\begin{equation}
e[n]=r[n]-\hat{s}_t[n]=r[n]-\textbf{w}^{T}\textbf{s}_r[n]\;.
\end{equation}
with $\textbf{w}$ being the equaliser coefficient vector and $\textbf{s}_r[n]$ holding the corresponding received signal samples from $s_r[n]$
\begin{eqnarray}
\textbf{w}&=&[w_0, w_1, \cdots, w_{L-1}]^T\nonumber\\
\textbf{s}_r[n]&=&[s_r[n], s_r[n-1], \cdots, s_r[n-L+1]]^{T}\;.
\end{eqnarray}

Following the derivations in \cite{liu14f,liu14p}, we have the gradient of $J[n]$ with respect to the coefficient vector as follows
\begin{equation}
\nabla_{\textbf{w}}J[n]=-\frac{1}{2}\textbf{s}_r[n]e^*[n],
\label{eqn:gradient}
\end{equation}
which leads to the following update equation for the coefficient vector with a step size of $\mu$, i.e. the QLMS algorithm:
\begin{equation}
\textbf{w}[n+1] = \textbf{w}[n]+\mu(e[n]\textbf{x}^{*}[n]).
\label{eq:update_weight_vector}
\end{equation}

For a solution equivalent to the classic Wiener filter in the complex-valued case, based on the instantaneous gradient result of \eqref{eqn:gradient}, the optimum solution $\textbf{w}_{opt}$ should satisfy
\begin{eqnarray}
E\{-\frac{1}{2}\textbf{s}_r[n]e^*[n]\}=0,
\label{eqn:gradient1}
\end{eqnarray}
i.e.
\begin{eqnarray}
E\{\textbf{s}_r[n]e^*[n]\}&=&E\{\textbf{s}_r[n]r^*[n]-\textbf{s}_{r}[n]\textbf{s}^H_r[n]\textbf{w}^*_{opt}\}\nonumber\\
&=&\textbf{p}-\textbf{R}_{s_r}\textbf{w}_{opt}^*=0\;,
\label{eqn:gradient2}
\end{eqnarray}
where $\textbf{p}=E\{\textbf{s}_r[n]r^*[n]\}$ and $\textbf{R}_{s_r}=E\{\textbf{s}_r[n]\textbf{s}^H_r[n]\}$.
Then we have
\begin{equation}
\textbf{w}^*_{opt}=\textbf{R}_{s_r}^{-1}\textbf{p}\;.
\end{equation}
We can use the above equation to obtain the optimum weight vector directly.

\begin{figure}
	\centering
	\includegraphics[scale=0.69]{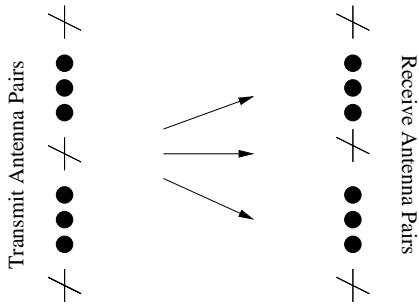}
	\caption{A MIMO system with multiple antenna pairs at both the transmitter and the receiver sides, where each pair is composed of two differently polarised antennas.}
	\label{fig:MIMOarray}
\end{figure}

We can extend this design to multiple antenna pairs for beamforming to suppress other quaternion-valued interfering signals, as shown in Fig.~\ref{fig:MIMOarray} for a general multiple-input-multiple-output (MIMO) system.

\section{Simulation Results}
\label{sec:sim}
In the following, we  give two sets of simulation results, one based on the structure in Fig.~\ref{fig:channelmodel} and on based on Fig.~\ref{fig:MIMOarray}.

\begin{figure}
	\centering
	\includegraphics[scale=0.49]{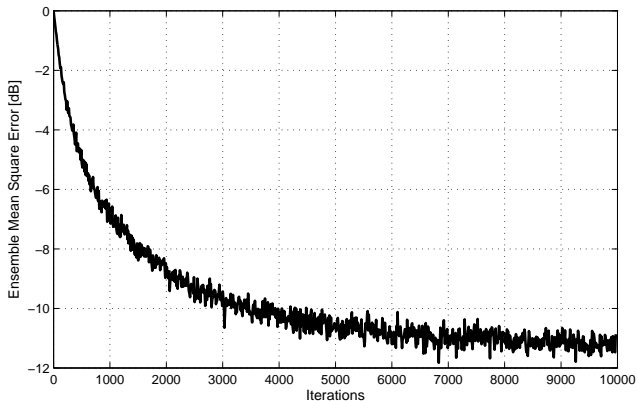}
	\caption{Learning curve of the quaternion-valued equaliser.}
	\label{fig:learningcurve}
\end{figure}

For the first set of simulations, the signal transmitted is 16-$\mbox{Q}^2\mbox{AM}$ modulated and the SNR at the receiver side is 20 dB with quaternion-valued Gaussian noise. The channel impulse response $f_c[n]$ is a 4-tap quaternion-valued FIR filter with a Gaussian-distributed coefficients value. The equaliser filter has a length of $15$. The learning curve based on averaging $200$ simulation runs is shown in Fig.~\ref{fig:learningcurve}, with about $-12$ dB error at the steady state, indicating a reasonable channel estimation result.

\begin{figure}
	\centering
	\includegraphics[scale=0.49]{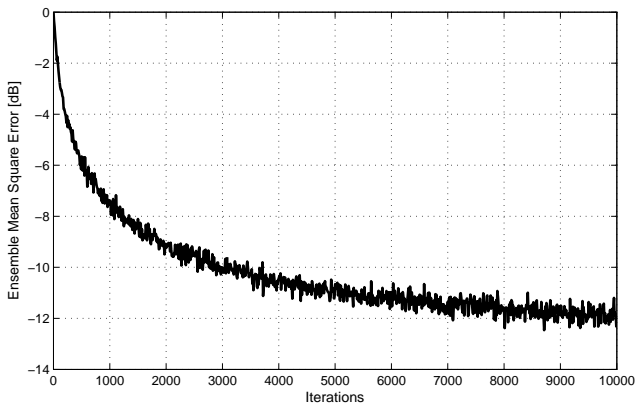}
	\caption{Learning curve of the quaternion-valued equaliser/MIMO beamformer.}
	\label{fig:learningMIMO}
\end{figure}

In the second set of simulations, we consider a $2\times2$ MIMO array and the two transmitted quaternion-valued signals have the same normalised power, with an SNR of 20 dB at the transmitter side. All the other parameters are the same as the first one. Fig.~\ref{fig:learningMIMO} shows the result, again a reasonable performance.

\section{Conclusions}
\label{sec:concl}

A 4-D modulation scheme using quaternion-valued representation based on two antennas with different polarisation directions has been studied for wireless communications. A quaternion-valued signal processing algorithm is employed for both equalisation and interference suppression/beamforming. Two sets of simulation results were provided to show that such a scheme can work effectively in both the single-input-single-output and multiple-input-multiple-output cases and therefore can be considered as a viable approach for future wireless communication systems.


\end{document}